\documentclass[letterpaper]{article}
\usepackage{aaai16}
\usepackage{times}
\usepackage{helvet}
\usepackage{courier}
\usepackage{setspace}
\usepackage{enumitem}
\usepackage{amsmath}
\usepackage{mathtools}
\usepackage{todonotes}
\usepackage{graphicx}
\usepackage{epstopdf}
\usepackage{wrapfig}

\usepackage{algorithmicx}
\usepackage{algpseudocode}
\usepackage{algorithm}

\usepackage{caption}
\usepackage{subcaption}
\usepackage{multirow}
\usepackage{xcolor}
\usepackage{framed}
\mathchardef\mhyphen="2D
\colorlet{shadecolor}{blue!20}
\usepackage{enumitem}
\setlist{nolistsep}
\setlist[description,0]{leftmargin=0pt}
\usepackage{setspace}
\newcommand{\norm}[1]{\left\lVert #1 \right\rVert}
\DeclareMathOperator*{\argmin}{arg\,min}
\DeclareMathOperator*{\argmax}{arg\,max}

\title{Predicting Social Status via Social Networks:\\ A Case Study on University, Occupation, and Region}

\author{
	Hao Fu\footnotemark[1], Xing Xie\footnotemark[2], Yong Rui\footnotemark[2], Defu Lian\footnotemark[3], Guangzhong Sun\footnotemark[1], Enhong Chen\footnotemark[1] \\
	\footnotemark[1]University of Science and Technology of China \\
	\footnotemark[2]Microsoft Research \\
	\footnotemark[3]University of Electronic Science and Technology of China
}
\begin{document}
\copyrighttext{This work was done when the first author was visiting Microsoft Research.}

\maketitle

\begin{abstract}
Social status refers to the relative position within the society. It is an important notion in sociology and related research. The problem of measuring social status has been studied for many years. Various indicators are proposed to assess social status of individuals, including educational attainment, occupation, and income/wealth. However, these indicators are sometimes difficult to collect or measure.

We investigate social networks for alternative measures of social status. Online activities expose certain traits of users in the real world. We are interested in how these activities are related to social status, and how social status can be predicted with social network data. To the best of our knowledge, this is the first study on connecting online activities with social status in reality.

In particular, we focus on the network structure of microblogs in this study. A user following another implies some kind of status. We cast the predicted social status of users to the ``status'' of real-world entities, e.g., universities, occupations, and regions, so that we can compare and validate predicted results with facts in the real world. We propose an efficient algorithm for this task and evaluate it on a dataset consisting of 3.4 million users from Sina Weibo. The result shows that it is possible to predict social status with reasonable accuracy using social network data. We also point out challenges and limitations of this approach, e.g., inconsistence between online popularity and real-world status for certain users. Our findings provide insights on analyzing online social status and future designs of ranking schemes for social networks.
\end{abstract}

%\category{H.2.8}{Database Management}{Database Applications}[Data mining]

%\keywords{Social status, social network, ranking, membership}

\section{Introduction}
\label{sec:introduction}
Social status is the relative position of a person or group within the society. It helps us to understand the stratification of society, and serves as reference for sociology and related research. Social status itself is an abstract concept. For many years, sociologists have discussed how to measure social status in a given society. Various indicators are proposed to assess social status, among which educational attainment, occupation, and income/wealth are the most commonly adopted \cite{adler2000relationship}. However, these indicators are expensive to collect at a large scale and sometimes difficult to measure. For example, to measure the socio-economic status of occupations, 73,901 respondents were surveyed in \cite{ganzeboom1996}.

% University ranking assesses the quality and reputation of higher education. Although being often criticized, it serves as an important criteria of the decision in college application. Social-economic status and prestige of occupations have been widely studied in sociology research. Occupations are usually ordered and assigned with status scores in these studies. Countries and regions are often compared in terms of economy and culture. In order to understand the difference, rankings of various statistics are employed. The rankings are sometimes difficult to obtain, since they require explicit statistics (e.g., the average income of a certain occupation or region) or implicit properties (e.g., the reputation of universities or occupations). For example, to measure the socio-economic status of occupations, over 73,901 respondents were surveyed in the work of Ganzeboom and Treiman \cite{ganzeboom1996}.

We are interested in alternative ways of deriving social status. The growing popularity of online social networks enables the possibility of measuring social status at a large scale. As users interact with each other in social networks, certain traits are exposed through their online activities, including social status. For instance, in a microblogging website, the number of followers is often used as an obvious indicator of status. Users with many followers are considered prestigious and influential \cite{hopcroft2011followback}. In previous works, structural stratification of the Twitter network \cite{hopcroft2011followback} and signed networks \cite{leskovec2010} has been observed. However, it is unclear how it is related to social status in reality. In this study, we focus on the impact of network structure on social status and leave other aspects (e.g., content of tweets) for future work.

We investigate the feasibility of predicting social status via online social networks in the following way. We develop an algorithm that is expected to predict social status, and then compare the results with commonly adopted indicators of social status, e.g., educational attainment, occupation, and income/wealth. If a consistent alignment is observed, we may presume that social status can be successfully predicted with social network data.

We consider educational attainment and occupation, but omit income/wealth in this study due to privacy issues in our dataset. Educational attainment can be measured with the highest degree \cite{adler2000relationship}, years in school, or the prestige of school that one is affiliated to or graduated from \cite{hevenstone2008academic}. We use the prestige of university in this study. Various measures of occupational status have been developed in the past decades. We leverage the results of \cite{ganzeboom1996} to measure occupational status. Additionally, we also consider the region that one lives in as an indicator of social status. As the development of regions within a country is usually unbalanced, it could yield difference of social status of residents.

For prediction, we follow the methodology of \cite{ganzeboom1996} where scores of respondents are aggregated to status scores of occupations. In our case, we first measure social status of individual users based on network structure, and then cast it to ``status'' of real-world entities, namely universities, occupations, and regions. An advantage of this approach is that one may easily compare the status scores of the entities with corresponding indicators and directly observe the difference.

We notice that people are the building blocks of the entities in question. A university can be regarded as a group of affiliated students and employees. An occupation can be represented by workers doing the same job. A region consists of residents who live in it. In many websites, users may specify various information about themselves, e.g., universities that they attended, current/former occupations, and regions that they live in. We denote such information as categorical attributes. Universities, occupations, and regions are represented as \textit{groups} of users. \textit{Members} of a group share the same attribute. We refer to such user-group relations as \textit{membership}. The notion of membership is an abstraction of user attributes, and should not be confused with terms in community detection literature. We use the term \textit{group status} to denote the generalization of social status of group members. It capture the similarity of its members in terms of social status. For simplicity, we represent it as a real-valued measure in this study. A high value indicates that the members have high social status. For example, professionals, when viewed as a whole group, are generally considered more prestige than people with other occupations. The measures of occupational status in \cite{ganzeboom1996} can be considered as an instance of group status.

A fundamental obstacle to this study is the noise and bias in social network data. This is deeply rooted in the nature of online social networks, where users can specify whatever they want about themselves. One may pretend to be a graduate of a famous university for better public recognition, or provide wrong information for privacy concern. Even if one can distinguish some users who honestly specified their profiles, it is still unclear if they represent the average population, or they are just those who try to promote themselves.

To obtain reliable profiles, we only consider \textit{verified} users who are required to provide proof of identity to the website. A drawback is that verified users are mostly influential and popular. Using only verified users would lead to biased results. We handle this issue by proposing a propagation algorithm that infers the membership of remaining users, expecting to enrich the set of reliable profiles. Although users can specify anything in their profiles, they have no direct control on the profiles of friends, which possibly reveal their real attributes \cite{mislove2010}.

We make two major contributions in this paper.
\begin{itemize}
	\item
	We propose a method that predicts group status based on social network structure, consisting of two steps. For the first step, we propose an efficient propagation algorithm to infer missing membership. For the second step, we discuss possible measures of social status of individuals and how group status is calculated.
	\item
	We evaluate our approach on a dataset consisting of 3.4 million users and three types of membership (university, occupation, and region). We compare the predicted status with various facts, and find good correlation between social network and social status in reality. However, inconsistence for certain types of users is also observed.
\end{itemize}

We first give an introduction of our dataset and a concrete definition of our problem in Section \ref{sec:preliminary}. The two steps of our method are described in Section \ref{sec:infer} and Section \ref{sec:ranking} respectively. We then evaluate our method and discuss the results in Section \ref{sec:exp}. Related works are discussed in Section \ref{sec:related-work}. A discussion of the findings is presented in Section \ref{sec:conclusion}.

\section{Preliminary}
\label{sec:preliminary}

\subsection{Dataset}
\label{sec:preliminary-dataset}
We collected our dataset from Sina Weibo, a popular microblogging website in China. We crawled our dataset in May, 2014 using the website's API. We applied the following strategy to obtain a reasonably ``good'' sample \cite{leskovec2006sample} from the website. We first made a random sample of tweets posted during April and May, 2014, from the public timeline, expecting to collect a uniform sample of active users. After removing duplicate authors, we collected 49,719 unique users as seeds. We crawled the 2-hop followees of the seed users to obtain a social graph.

We then crawled profiles of the users, among which $8.7\%$ users were verified. $52.6\%$ of the verified users were verified as individuals, while the others were non-individuals, e.g., government, enterprise, and news media. We removed non-individual users since we study social status of real human. Finally, we obtained a graph with 3.4 million nodes and 475 million directed edges.

The website's API provides very limited data of users. It does not contain any information about educational attainment or occupation, though they are accessible via user's profile page. We further crawled the profile pages of verified individual users. Due to the access limit of the website, we were only able to crawl 63,346 valid pages, covering $39\%$ of the verified individual users. We extracted the membership from profiles and pages as follows.
\begin{description}
	\item[University]
	Users can specify the universities that they attend or graduate from. A university may have multiple synonymous names. For example, users from Peking University may specify ``Peking U.'' or simply ``PKU''. We manually inspected all distinct university names and replaced them with full names, so that synonymous names were mapped to the same university. We discarded universities having less than 20 members, because inference on such small samples is unreliable. We ended up with 10,256 users who specified their universities. 158 distinct universities were involved, most of which are located in China.
	
	\item[Occupation]
	Users can specify their occupational titles in profiles. An occupation may have multiple synonymous titles, e.g., programmer and software engineer. In addition, the occupational titles are sometimes ambiguous. One may specify ``engineer'', but it is unclear if it means software engineer, mechanical engineer, or something else. To solve this problem, we referred to the International Standard Classification of Occupations (ISCO) \cite{ganzeboom1996} to unify the occupational titles. ISCO is a hierarchical classification system mostly derived from skill requirements at the expense of industry distinctions. It contains 9 major groups and three further levels: 28 sub major groups, 116 minor groups, and 290 unit groups.
	
	We inspected the top 1,000 frequent occupational titles in our dataset, and manually mapped them to ISCO groups. Ambiguous occupational titles were difficult to be mapped to fine-grained levels, so we focus on the first level of ISCO (9 groups). The 1,000 occupational titles were classified to 7 major groups. We ended up with 10,582 users that were associated to ISCO groups.
	
	\item[Region]
	Sina Weibo allows users to specify the regions that they live in. The regions are organized according to the administrative divisions of China, including 4 cities, 28 provinces, and 2 special administrative regions. We followed this division and ignored oversea users. We obtained 72,252 known users spreading in the 34 regions.
\end{description}

We are concerned about the sampling bias of our dataset. For each of the three types of membership, over 84\% users are associated to only one group. We find that the distribution of users in these groups is unbalanced. Professional occupations account for 32\% users. About half users live in two major cities, Beijing and Shanghai. We checked with the website's statistics using its filtered search, and find the distributions are roughly consistent, so we believe our dataset is a representative sample of the website.

\subsection{Problem Formulation}
\label{sec:preliminary-problem}

We model a microblogging website as a directed graph $G = (V, E)$. Each node in $V$ corresponds to a user in the website. For simplicity, we use the terms \textit{user} and \textit{node} interchangeably. A directed edge $(u, v)$ is included in $E$ if and only if user $u$ follows user $v$. We refer to the user being followed as a \textit{followee} and the other user as a \textit{follower}. If two users are following each other reciprocally, they are called \textit{friends}. We represent the sets of followees, followers, and friends of $u$ as $N_O(u)$, $N_I(u)$, and $N_R(u)$ respectively.

We denote $H = \{h_1, h_2, ..., h_m\}$ as the set of groups, where $m$ is the number of groups. For example, $H$ may indicate all the distinct universities in the dataset. The membership of a user $u$ is denoted as $H_u$, which is a subset of $H$. User $u$ is a member of group $h_k$, if $h_k \in H_u$. Users with known membership are denoted as $V^+$, and unknown users are denoted as $V^-$. This definition captures the case that a user is associated to multiple groups.

We formulate the prediction problem as follows. We first calculate the status score $P_u$ for each $u \in V$. Given membership of partial known users, we then compute the group status score $\pi_i$ for each $h_i \in H$. $P_u$ and $\pi_i$ assess the social status of users and groups. High values indicate high status. The group status is expected to align with the indicators of social status in reality.

\section{Membership Inference}
\label{sec:infer}
As the first step, we introduce how we infer unknown membership from verified users. We start by observing homophily in the dataset. Based on the observation, we propose a supervised propagation algorithm. We then discuss how the model is trained efficiently.

\subsection{Homophily}
\label{sec:infer-homophily}
In social networks, homophily refers to the property that users with similar characteristics tend to be associated with each other. Previous works have discovered that users are more likely to follow each other reciprocally if they share similar topics \cite{weng2010wsdm} or friends \cite{hopcroft2011followback}. We wonder if members of the same group are likely to follow each other.

Given a pair of users who follow each other reciprocally, we calculate the probability that they are members of the same group. We also calculate this probability when there is only a one-way ``follow'' between the two users, as well as when they do not follow each other. The result (Table \ref{tab:prob-same-attr}) shows that connected users are more probable to be in the same group than those disconnected. This tendency is more obvious if the users follow each other reciprocally. This observation confirms the homophily and reveals the relation between graph structure and membership.

\begin{table}
\caption{Probability of being members of the same group. The standard error of estimated probability is less than $10^{-4}$.}
\label{tab:prob-same-attr}
\centering
\begin{tabular}{cccc}
\hline
\textbf{User pair} & \textbf{University} & \textbf{Occupation} & \textbf{Region} \\
\hline
Reciprocal & \textbf{0.135} & \textbf{0.483} & \textbf{0.530} \\
\hline
One-way & 0.075 & 0.468 & 0.381 \\
\hline
Disconnected & 0.018 & 0.287 & 0.240 \\
\hline
\end{tabular}
\end{table}

\subsection{Propagation}
\label{sec:infer-propagate}

We represent the strength of membership of an unknown user $u$ as $Q_u = (Q_{u1}, Q_{u2}, ..., Q_{um})$. $Q_{ui}$ denotes the probability of $u$ being a member of group $h_i$. The membership of a known user $v$ is represented as $Q^0_v$ in a similar way that
\begin{equation}
Q^0_{vi} =
\begin{cases}
1 & h_i \in H_v \\
0 & h_i \notin H_v
\end{cases}
\nonumber
\end{equation}

According to homophily, one may infer a user as a member of the most probable group that her friends belong to. By regarding the newly inferred users as known, the membership of more users can be inferred. This suggests a propagation algorithm for membership inference. Various efficient propagation algorithms have already been proposed for similar tasks \cite{raghavan2007,pennacchiotti2011}. However, little attention has been paid to the strength of social ties. One may share a lot in common with a close friend but know nothing about an acquaintance.

In recent years, a family of supervised random walk algorithms have been proposed to improve the quality of link prediction \cite{backstrom2011rw} and web page ranking \cite{gao2011srv}. An edge weight function is learned to bias the random walk. We leverage the techniques developed for supervised random walk to learn social tie strength efficiently, aiming to improve the accuracy of membership inference in large-scale networks.

In our scenario, we consider the social tie between two users is strong if they share the same membership. A close friend should have more impact on the inference. We denote the social tie strength between $u$ and $v$ as a non-negative function $f(u, v)$. Note that $f(u, v)$ is not symmetrical, i.e.,  $f(u, v) \neq f(v, u)$. While $u$ regards $v$ as a good friend, $v$ may not treat $u$ in the same way. This leads to a weighted propagation:
\begin{equation}
\label{eqn:weighted-propagate}
\hat{Q}_u = \frac{\sum_{v \in N_R(u)} f(u, v) Q_v}{\sum_{v \in N_R(u)} f(u, v)}
\end{equation}
We calculate $\hat{Q}_u$ for all $u \in V^-$, and replace $Q_u$ with the newly calculated $\hat{Q}_u$. This process is repeated until all $Q_u$ converge. $Q_u$ is initially assigned with $1 / 2$ for $u \in V^-$. For known users $V^+$, we fix $Q_u = Q^0_u$, so that their membership is propagated to their friends repeatedly.

\subsection{The Model}
\label{sec:infer-model}
We assume that $f(u, v)$ depends on the features $X_{uv} = (x^{uv}_1, x^{uv}_2, ..., x^{uv}_k)$. In our experiments, we used 11 features, which are the number of followers, the number followees, the number of friends, PageRank, and reversed PageRank (reversing the direction of edges) of both users, as well as the number of common friends. We omit the analysis of features due to space limit. We follow the definition in \cite{backstrom2011rw} and define the strength $f(u, v)$ as a function of the features:
\begin{equation}
\label{eqn:propagate-strength}
f(u, v) = \frac{1}{1 + \exp \{ - X_{uv} \cdot \mathbf{w} \}}
\end{equation}
While other forms of $f(u, v)$ are possible, we find this definition works well in our experiments.

We determine the parameter $\mathbf{w} = (w_1, w_2, ..., w_k)$ from the dataset. We randomly split the known users $V^+$ into two sets, namely \textit{seed nodes} $V_S$ and \textit{target nodes} $V_T$. We start with some random value of $\textbf{w}$. We pretend that the membership of target nodes $V_T$ is unknown. Equation (\ref{eqn:weighted-propagate}) is applied iteratively to infer the membership of unknown nodes and target nodes. We then check the estimation error between the inferred membership $\hat{Q}_u$ and the actual membership $Q^0_u$ of target nodes, and update $\mathbf{w}$ so that the error is reduced. The above process is repeated until the estimation error is minimized. This idea is formulated as an optimization problem that minimizes the loss function
\begin{equation}
\label{eqn:propagate-loss}
L(\mathbf{w}) = \frac{1}{2}\sum_{u \in V_T} \norm{\hat{Q}_u - Q^0_u}^2 + \frac{\mu}{2} \norm{\mathbf{w}}^2
\end{equation}
A regulation term is included to avoid overfitting. We leverage the techniques in \cite{gao2011srv} to solve the optimization problem efficiently. Note that the propagation described by Equation (\ref{eqn:weighted-propagate}) is equivalent to
\begin{equation}
\argmin_Q \sum_{u \notin V_T} \norm{\hat{Q}_u - Q_u}^2
\nonumber
\end{equation}
A new loss function is obtained by including the above term
\begin{equation}
\label{eqn:propagate-loss-alter}
\begin{split}
L(\mathbf{w}, Q) = & \frac{1}{2}\sum_{u \in V_T} \norm{\hat{Q}_u - Q^0_u}^2 + \\
& \frac{\lambda}{2} \sum_{u \notin V_T} \norm{\hat{Q}_u - Q_u}^2 + \frac{\mu}{2} \norm{\mathbf{w}}^2
\end{split}
\end{equation}

Now that the membership strength $Q$ is treated as an argument of the loss function. It is determined by solving the optimization problem rather than propagating using Equation (\ref{eqn:weighted-propagate}). We apply a stochastic gradient descent algorithm to optimize $L(\mathbf{w}, Q)$. $Q$ and $\mathbf{w}$ are updated simultaneously. With a single server, it takes about 40 iterations in less than 1 hour to converge on our dataset. We then obtain the membership strength $Q$ immediately.

\section{Predicting Status}
\label{sec:ranking}

Now we proceed to predict group status based on the inferred membership. We first discuss measures of social status of users. We then discuss how group status is predicted.

\subsection{Measuring Status}
\label{sec:ranking-user}

A theory of status in social networks was implicit in \cite{guha2004} and later developed in \cite{leskovec2010}. The status theory assumes a status ordering of users and that directed links imply the status ordering. In a microblogging website, a link $(u, v)$ can be explained as $u$ having a lower status than $v$ has. If $u$ regards $v$ as having a higher status, while $w$ is considered more prestige than $v$, the status theory predicts that $w$'s status is also higher than $u$'s.

We examine if the Weibo network can be explained by the status theory. The status theory makes prediction on the type of triangles. Given that $(u, v)$ and $(v, w)$ both exist, it predicts that the link between $u$ and $w$ must be $(u, w)$ rather than $(w, u)$. We search for evidence by counting the number of triangles. A triangle of type I consists of $(u, v)$, $(v, w)$, and $(u, w)$, which is a positive evidence of the status theory. A triangle of type II is a directed circle consisting of links through $u$, $v$, and $w$, which is a negative evidence. It can be shown that every triangle that contains exactly three links is isomorphic to either of the two triangles. In our dataset, the ratio between the two types is 5.8, which means about five times more triangles of type I is presented in the network than type II. This clearly shows that the Weibo network can be well explained by the status theory.

In \cite{leskovec2010}, a heuristic was introduced to order users from high status to low status. It starts with a random ordering of nodes, and swap the positions of randomly selected nodes repeatedly, aiming to increase the number of links from low status to high status. In a graph with millions of nodes, we find it rather time-consuming to find a good solution, so we consider other measures that can be efficiently computed, including the number of followers, eigenvector centrality, and PageRank. It is worthwhile to investigate and design more complicated measures, but it is beyond the scope of this paper. These measures follow the similar intuition that a user is more prestige than her followers. Note that these network-based measures are not measures of social status yet, until we validate with indicators of social status in reality. 

The number of followers is a natural and obvious measure of a user's position in the social network. A user with millions of followers has direct influence on that many users, so one may assume that she has a high social status. However, this measure does not accurately capture the status, e.g., it can be easily manipulated \cite{ghosh2012www}.

Eigenvector centrality \cite{bonacich1987power} measures the importance of a user in a social network. It improves the number of followers by considering the structure of the whole network. It is defined based on the concept that a user is as important as those who she has influence on. Specifically, the eigenvector centrality of a user is proportional to the sum of eigenvector centrality of her followers. Let $A$ be the adjacency matrix of the graph $G = (V, E)$, i.e., $A_{uv} = 1$ if $(u, v) \in E$, or $A_{uv} = 0$ otherwise. Eigenvector centrality is essentially the principal eigenvector of $A^T$. Given that a normal user is only capable to read a limited number of tweets per day, a user having too many followees is unlikely to be influenced. Such users do not contribute much to the influence of their followees. A limitation of eigenvector centrality is that it does not consider such cases.

PageRank has been widely used to rank web pages. It is adopted for ranking users in social networks \cite{tang2009influence,weng2010wsdm}. It improves eigenvector centrality by weighing nodes with degrees. Users following many others contribute less to their followees. PageRank is defined as the stationary distribution of a random walk process:
\begin{equation}
\label{eqn:pagerank}
P_v = d \cdot \sum_{u \in N_I(v)} \frac{P_u}{|N_O(u)|} + \frac{1 - d}{|V|}
\end{equation}
The damping factor $d$ is usually assigned with $0.85$. PageRank is shown to be a good measure of status \cite{hopcroft2011followback} and influence \cite{weng2010wsdm} in microblogs, so we focus on it in this paper.

\subsection{Group Status}
\label{sec:ranking-group}
Now we discuss how status score $P$ of users is transformed to group status score $\pi$. A natural assumption is that a group's status is determined only by its members. Two groups may interact with each other via the links between their members, but such interactions were supposed to be captured by $P$. Recall that we represent the membership of users as a set of probabilities $Q$. If we know $u$ as a member of $h_i$ for sure, i.e., $Q_{ui} = 1$, $\pi_i$ should be affected by $P_u$ to the maximum extent. If we are uncertain, e.g., $Q_{ui}=1/2$, less impact to $\pi_i$ should be expected.

Another assumption is that a group's status should be irrelevant to its size. First, the number of members is not a dominating factor of the status of a group. For example, most university rankings focus on the quality and reputation of higher education, rather than the number of students and employees. Second, membership of only a sample of users is known in this study. Even with the membership inference, we do not know for sure the exact number of actual members of a group. Third, the number of active social network users may vary among groups. Groups with more active users may get higher status scores. By canceling the number of social network users, we may partially discard such bias.

Putting the above assumptions together, we define a group's status score as the ratio between the expected sum of its members' status score and its expected size, i.e.,
\begin{equation}
\label{eqn:group-status}
\pi_i = \frac{\sum_{v \in V} Q_{vi} P_v}{\sum_{v \in V} Q_{vi}}
\end{equation}
The contribution to $\pi_i$ of a user $u$ is proportional to $u$'s membership strength. A ranking can be generated by ordering groups in descending order of $\pi_i$.

\section{Experiments}
\label{sec:exp}

\begin{table}
	\caption{Accuracy of membership inference}
	\label{tab:avg-accuracy-membership}
	\centering
	\begin{tabular}{cccc}
		\hline
		& Method & Accuracy & Avg.\ Accuracy \\
		\hline
		\multirow{4}{*}{\textbf{University}} & SP & 0.245 & \textbf{0.199} \\
		& UP & \textbf{0.246} & 0.135 \\
		& LP & 0.086 & 0.015 \\
		& Hybrid & 0.124 & 0.094 \\
		\hline
		\multirow{4}{*}{\textbf{Occupation}} & SP & \textbf{0.399} & \textbf{0.287} \\
		& UP & \textbf{0.399} & 0.210 \\
		& LP & 0.392 & 0.187 \\
		& Hybrid & 0.385 & 0.252 \\
		\hline
		\multirow{4}{*}{\textbf{Region}} & SP & \textbf{0.760} & \textbf{0.522} \\
		& UP & 0.643 & 0.321 \\
		& LP & 0.531 & 0.151 \\
		& Hybrid & 0.665 & 0.477 \\
		\hline
	\end{tabular}
\end{table}

In this section, we evaluate the performance of the proposed membership inference algorithm. And then we compare and discuss the predicted group status with reality.

\subsection{Membership Inference}
\label{sec:exp-membership}
\subsubsection{Settings}
Membership inference estimates unknown membership from known users. Given a type of membership, we split the known users into two sets, namely training set and testing set. We feed our algorithm with the training set, and evaluate the accuracy of inferred membership on the testing set. We consider the most probable group ($\argmax_{h_i \in H} Q_{ui}$) as the inferred group. The inference is considered correct if the user is an actual member of the inferred group. We define the accuracy as the fraction of correctly inferred users. As the group size is unbalanced, we also consider a balanced version of accuracy, which is the average over the accuracy on each group.

Our approach, supervised propagation (SP), requires seed nodes and target nodes. We keep 80\% of the training set for seed nodes, and leave the other for target nodes. The parameter $\lambda$ trades off between the inference error on target nodes and the convergence of propagation. The parameter $\mu$ controls the extent of overfitting. We find that setting $\lambda = 1$ and $\mu = 1$ works well in most cases. We repeat the iteration until the relative improvement of $L(\mathbf{w}, Q)$ is less than $1\%$. We also evaluated the following approaches for comparison.

\begin{description}
\item[Uniform Propagation (UP)]
To investigate the effect of the learned social tie strength $f(u, v)$, we consider a simplified model where $f(u, v)$ is fixed to a constant for each $(u, v) \in E$, so that the membership is propagated uniformly.
\item[Label Propagation (LP)]
A clustering algorithm described in \cite{raghavan2007} starts with a set of known nodes and update the label of other nodes iteratively. Every node takes the most frequent label in its neighbors. This algorithm can be considered as a ``hard'' version of UP, as membership is indicated with categorical labels rather than probabilities.
\item[Hybrid]
Pennacchiotti and Popescu (\citeyear{pennacchiotti2011}) proposed a hybrid algorithm that infers binary labels on Twitter. They first apply a machine learning algorithm to infer node labels individually. The inferred label is then propagated via edges to correct previous errors. The original algorithm can only handle binary attributes. We decompose the membership of $m$ groups into $m$ binary attributes, and apply the algorithm separately. To make a fair comparison, only profile features and social network features are used.
\end{description}

We also considered other recent works \cite{backstrom2010,mislove2010,zhao2013,li2014} that might be used for membership inference. A discussion of them is presented in Section \ref{sec:related-work}.

\subsubsection{Results}

We use a 10-fold cross validation in the experiments. The results are reported as the averages (Table \ref{tab:avg-accuracy-membership}). UP performs better than LP, indicating the advantage of representing membership as probabilities instead of discrete labels. Hybrid performs quite well on inferring occupations and regions, but not as good as SP. An explanation is that structural information is not fully utilized in Hybrid, since it propagates labels only to 1-hop neighbors. SP achieves the highest accuracy on the three types of membership. This shows the power of the learned social tie strength $f(u, v)$. We omit detailed discussion of the results here, since it is not the main focus of this paper.

\subsection{Prediction}
\label{sec:exp-rank}

\begin{table}
	\caption{AUC of predicting top universities}
	\label{tab:predict-university}
	\centering
	\begin{tabular}{cccc}
		\hline
		\textbf{Method} & \hspace*{-1em} \textbf{Project 211} & \textbf{Project 985} & \textbf{C9 League} \\
		\hline
		SP & \textbf{0.612} & \textbf{0.770} & \textbf{0.903} \\
		\hline
		UP & 0.601 & 0.767 & 0.882 \\
		\hline
		PR & 0.431 & 0.618 & 0.776 \\
		\hline
	\end{tabular}
\end{table}

Now we have the inferred membership strength $Q$. We first calculate status scores $P$ of users, and then put them together to obtain group status scores $\pi$ as described in Section \ref{sec:ranking}. Membership inference is expected recover missing membership so that we can have better prediction of group status. Our first concern is about its ability to improve the quality of prediction. Our second concern is that whether the predicted group status is consistent with social status in the real world. If not, when and why?

For the first concern, we evaluate approaches with different settings of membership inference. We compare the results with facts collected from reality across domains. For the second concern, we distinguish cases when inconsistence is observed and conduct analysis on possible factors. The following approaches are evaluated.
\begin{description}
	\item[SP]
	This is the major approach proposed in this paper, where missing membership is inferred with the supervised propagation algorithm.
	\item[UP]
	We consider a variant of our method in which uniform propagation is used for membership inference. We compare with this approach to validate that improving membership inference has a positive effect on predicting group status.
	\item[PR]
	In this approach, we do not perform membership inference. We calculate PageRank on the whole network and then take a group's status score as the average of its known (verified) members' PageRank. We evaluate this approach to verify the necessity of membership inference.
\end{description}

Evaluation is performed with membership of university, occupation, and region. We discuss the results separately.

\subsubsection{University}
\label{sec:exp-rank-university}

\newcommand{\cpbox}[2]{\parbox[c]{1.5in}{\scriptsize \vspace{0.2em} \textsuperscript{#1}#2 \vspace{0.1em}} \hspace*{-10em} }

\newcommand{\rkbox}[1]{\hspace*{-0.5em}#1\hspace*{-1em}}

\begin{table}
	\caption{Top 20 universities by group status}
	\label{tab:rank-university}
	\centering
	\scriptsize
	\begin{tabular}{lll}
		\hline
		\rkbox{\textbf{\#}} & \parbox{1.43in}{\centering \textbf{SP}} & \parbox{1.43in}{\centering \textbf{PR}} \\
		\hline
		\rkbox{1} & \cpbox{123}{Peking Univ.} & \cpbox{}{Cheung Kong Graduate School of Business} \\
		\hline
		\rkbox{2} & \cpbox{1*}{Communication Univ. of China} & \cpbox{}{Southwest Univ. of Political Sci. and Law} \\
		\hline
		\rkbox{3} & \cpbox{*}{Central Academy of Drama} & \cpbox{*}{Shanghai Theatre Academy} \\
		\hline
		\rkbox{4} & \cpbox{123}{Tsinghua Univ.} & \cpbox{1}{China Univ. of Political Sci. and Law} \\
		\hline
		\rkbox{5} & \cpbox{*}{Shanghai Theatre Academy} & \cpbox{*}{P.L.A. Arts College} \\
		\hline
		\rkbox{6} & \cpbox{*}{Beijing Film Academy} & \cpbox{*}{Central Academy of Drama} \\
		\hline
		\rkbox{7} & \cpbox{12}{Renmin Univ. of China} & \cpbox{}{Univ. of International Relations} \\
		\hline
		\rkbox{8} & \cpbox{}{Tianjin Normal Univ.} & \cpbox{1}{Hunan Normal Univ.} \\
		\hline
		\rkbox{9} & \cpbox{123}{Fudan Univ.} & \cpbox{*}{Sichuan Fine Arts Institute} \\
		\hline
		\rkbox{10} & \cpbox{1}{China Univ. of Political Sci. and Law} & \cpbox{}{China Europe Int'l Business School} \\
		\hline
		\rkbox{11} & \cpbox{12}{Beijing Normal Univ.} & \cpbox{123}{Peking Univ.} \\
		\hline
		\rkbox{12} & \cpbox{12}{Huazhong Univ. of Sci. and Tech.} & \cpbox{1}{Univ. of Sci. and Tech. Beijing} \\
		\hline
		\rkbox{13} & \cpbox{123}{Nanjing Univ.} & \cpbox{*}{Beijing Film Academy} \\
		\hline
		\rkbox{14} & \cpbox{12}{East China Normal Univ.} & \cpbox{}{Tianjin Normal Univ.} \\
		\hline
		\rkbox{15} & \cpbox{123}{Shanghai Jiao Tong Univ.} & \cpbox{}{Guangdong Univ. of Tech.} \\
		\hline
		\rkbox{16} & \cpbox{123}{Zhejiang Univ.} & \cpbox{1*}{Central Conservatory of Music} \\
		\hline
		\rkbox{17} & \cpbox{12}{Nankai Univ.} & \cpbox{123}{Fudan Univ.} \\
		\hline
		\rkbox{18} & \cpbox{}{China Europe Int'l Business School} & \cpbox{}{Capital Normal Univ.} \\
		\hline
		\rkbox{19} & \cpbox{123}{Univ. of Sci. and Tech. of China} & \cpbox{}{Chinese Academy of Social Sciences} \\
		\hline
		\rkbox{20} & \cpbox{}{Southwest Univ. of Political Sci. and Law} & \cpbox{12}{Nankai Univ.} \\
		\hline
		\multicolumn{3}{l}{\scriptsize \vspace{0.1em} \hspace*{-1em} \textsuperscript{1}Project 211, \textsuperscript{2}Project 985, \textsuperscript{3}C9 League, \textsuperscript{*}Art/music/acting school}
	\end{tabular}
\end{table}

To assess the quality of predicted university status, we compare it with public rankings by third-parties. Multiple Chinese university rankings are published by different institutes, e.g., the ARWU rankings and the QS World University Rankings. These rankings focus on different factors of quality and reputation of university. However, their source data and criteria are rarely published, rendering them less convincing. Therefore, we refer to some kind of ``official'' classification of top universities. We consider Project 211, Project 985, and the C9 League. Project 211 and Project 985 are both Chinese government-run education projects\footnote{http://www.moe.gov.cn/}. Universities designated as Project 211 or Project 985 institutes are supported with first priority. Until 2015, 116 universities and 39 universities are sponsored by the two projects respectively. The C9 League consists of the first nine universities selected for Project 985. It is often referred to as the Chinese equivalent of the US Ivy League. These universities are generally regarded as top universities in China.

\newcommand{\optbox}[1]{\parbox[c]{1.53in}{\scriptsize \vspace{0.2em} \centering #1 \vspace{0.1em}} }
\newcommand{\opbox}[1]{\parbox[c]{1.6in}{\scriptsize \vspace{0.2em}#1\vspace{0.1em}}\hspace*{-10em}}
\begin{table*}
	\caption{Ranking of occupations by SIOPS, ISEI, and group status}
	\label{tab:rank-occupation}
	\centering
	\scriptsize
	\begin{tabular}{lllll}
		\hline
		\rkbox{\textbf{\#}} & \optbox{\textbf{SIOPS}} & \optbox{\textbf{ISEI}} & \optbox{\textbf{SP/UP}} & \optbox{\textbf{PR}} \\
		\hline
		\rkbox{1} & \opbox{Professionals} & \opbox{Professionals} & \opbox{Legislators, Senior Officials \& Managers} & \opbox{Legislators, Senior Officials \& Managers} \\
		\hline
		\rkbox{2} & \opbox{Legislators, Senior Officials \& Managers} & \opbox{Legislators, Senior Officials \& Managers} & \opbox{Professionals} & \opbox{Professionals} \\
		\hline
		\rkbox{3} & \opbox{Technicians \& Associate Professionals} & \opbox{Technicians \& Associate Professionals} & \opbox{Technicians \& Associate Professionals} & \opbox{Technicians \& Associate Professionals} \\
		\hline
		\rkbox{4} & \opbox{Craft \& Related Trades Workers} & \opbox{Clerks} & \opbox{Service, Shop \& Market Sales Workers} & \opbox{Skilled Agricultural \& Fishery Workers} \\
		\hline
		\rkbox{5} & \opbox{Clerks} & \opbox{Service, Shop \& Market Sales Workers} & \opbox{Clerks} & \opbox{Service, Shop \& Market Sales Workers} \\
		\hline
		\rkbox{6} & \opbox{Skilled Agricultural \& Fishery Workers} & \opbox{Craft \& Related Trades Workers} & \opbox{Craft \& Related Trades Workers} & \opbox{Clerks} \\
		\hline
		\rkbox{7} & \opbox{Service, Shop \& Market Sales Workers} & \opbox{Skilled Agricultural \& Fishery Workers} & \opbox{Skilled Agricultural \& Fishery Workers} & \opbox{Craft \& Related Trades Workers} \\
		\hline
	\end{tabular}
\end{table*}

We are aware that university ranking is a sensitive topic. A university can argue against a ranking because they are ranked lower than they think. To avoid this, we formulate a binary classification problem instead. Given a university, the task is to predict if it is one of the top universities (Project 211, Project 985, or the C9 League). We do this by sorting universities in descending order of group status score.

We adopt the standard notion of true positive rate and false positive rate to assess the performance. The true positive rate (TPR) is defined as the fraction of correctly identified top universities out of the actual top universities. The false positive rate (FPR) is defined as the fraction of misclassified non-top universities out of the actual non-top universities. The trade-off between TPR and FPR is visualized by the ROC curve. The overall performance is quantified by the area under the curve (AUC).

The performance of prediction is presented in Table \ref{tab:predict-university}. C9 League universities can be better predicted than Project 211 and Project 985 universities by the three approaches. In general, the results show positive correlation between the predicted group status and their real status.

The PR approach, which simply takes the average of known users' PageRank without membership inference, performs poorly. It is even worse than random ($\mathrm{AUC} < 0.5$) for Project 211 universities. We find that users from non-top universities are less likely to have their accounts verified. For the three classifications, the average number of verified users from non-top universities is less than 58, while the similar average among top universities is greater than 84. To understand this behavior, we manually inspect user profiles from top-ranked universities by PR. We find a tendency that graduates of top universities prefer to have their accounts verified for better public recognition, while only elite graduates of non-top universities would verify their accounts.  In other words, the PR approach compares average graduates of top universities with elite graduates of non-top universities, which does not make much sense.

This confirms our concern about the noise and bias in online data. Although we can obtain reliable profiles by considering only verified users, they turn out to be a biased sample of the total population. Certain methods are required to filter and clean the data. In our case, we propose a propagation algorithm to infer missing membership. The simplified version of the algorithm (UP) improves the performance of prediction significantly, implying that membership inference is an effective way to recover missing data. By learning the social tie strength (SP), the performance is further improved. Although the algorithm can not infer missing membership perfectly (Table \ref{tab:avg-accuracy-membership}), it turns out to be sufficient to predict group status reasonably well.

We present the top 20 universities by SP and PR in Table \ref{tab:rank-university}. A significant number of art, music, or acting schools are in the top 20. First, artists, musicians, and actors are actually more famous to the general public than other occupations, e.g., most people know Lady Gaga but rarely hear of Paul Erd\H{o}s. As most of them graduate from art, music, or acting schools, the status of these schools is promoted on social networks. Second, some law schools and business schools are inferred with high status. This can be explained by the occupational status of their graduates, e.g., lawyers and managers. The remaining universities by SP are mostly associated to Project 211, Project 985, or C9 League, while only 7 universities by PR are associated to Project 211.

In summary, we observe sampling bias of verified users, which leads to undesirable prediction of simple approaches (e.g., PR). By inferring missing membership, this issue can be partially corrected. Our result shows that it is possible to prediction status of universities with reasonable accuracy. However, inconsistence is observed for certain types of institutions. Their graduates gain extraordinary occupational influence on social networks by nature, and it is a stubborn problem on predicting social status via social networks.

\subsubsection{Occupation}
\label{sec:exp-rank-occupation}

In sociology, three indicators are commonly used for occupational status, namely occupational prestige, socio-economic index, and Erikson-Goldthorpe-Portocarero (EGP) class categories. Occupational prestige is a scalar measure of the respect with which an occupation is regarded by others. Socio-economic index scores are calculated as weighted averages of standardized measures of income and educational attainment. EGP class categories are nominal typologies that combine occupational information. 

Ganzeboom and Treiman \cite{ganzeboom1996} developed internationally comparable measures of the above three indicators, namely Standard International Occupational Prestige Scale (SIOPS), International Socio-Economic Index of Occupational Status (ISEI), and EGP class categories. As EGP class categories are not ordered, we only consider the other two here. We compare the above indicators with predicted status.

We order the 7 occupation groups from high status to low status in Table \ref{tab:rank-occupation}. The orderings by SP and UP are the same. In general, ISEI is more correlated to predicted group status than SIOPS, indicating that social status on Sina Weibo appears more relevant to income and educational attainment. Further investigation is required to justify this observation.

SP/UP and ISEI are similar, except that two pairs of adjacent groups are swapped. First, ``Professionals'' and ``Legislators, Senior Officials \& Managers'' are swapped. We inspected the occupational titles of the two groups. Scientists and actors are ranked among the top, but they consist of only 18\% professionals. The other group contains only corporate managers. We did not find any legislator or official in the top 1,000 frequent occupational titles. This is possibly due to the sensitivity of their identities. Corporate managers appear to be more popular than other professionals, e.g., journalists, doctors, and engineers. Second, ``Clerks'' and ``Service, Shop \& Market Sales Workers'' are swapped. Marketing related occupations, e.g., advertising, salesman, and public relation, consist of 85\% of the second group. We find evidence of link farming aiming to increase their influence. One method is following back anyone who follows them. Ghosh et.\ al \cite{ghosh2012www} find it an effective way to promote the ranking on Twitter. Marketers follow back 37\% of their followers on average, while the similar ratio is 24\% and 31\% for ``Clerks'' and all users. 

The above observations suggest that the slight difference between SP/UP and ISEI is mainly caused by the unbalanced distribution of known occupations as well as link farming by marketers. The first issue could be solved by considering fine-grained occupational groups, if we have more specified occupational titles in user profiles. For the second issue, anti-spam techniques could be helpful to discard the effect of link farming \cite{ghosh2012www}. The PR approach is less relevant to ISEI, suggesting the effect of membership inference.

\subsubsection{Region}
\label{sec:exp-rank-region}

\begin{table}
	\caption{Spearman correlation between rankings of regions by group status and social-economic statistics}
	\label{tab:rank-region-correlation}
	\centering
	\begin{tabular}{ccc}
		\hline
		\textbf{Method} & \textbf{Per capita income} & \textbf{Per capita GDP} \\
		\hline
		SP & 0.683 & 0.498 \\
		\hline
		UP & \textbf{0.690} & \textbf{0.502} \\
		\hline
		PR & 0.467 & 0.397 \\
		\hline
	\end{tabular}
\end{table}

Recall that we predict group status according to status of individuals. Correspondingly, we focus on per capita socio-economic statistics, namely per capita income and per capita GDP in 2014\footnote{http://www.stats.gov.cn/}, which are indicators of average income and wealth of citizens in each region.

We rank the regions according to the statistics and predicted group status. We measure the correlation with Spearman correlation coefficient. A perfect Spearman correlation of +1 or -1 occurs when the two rankings are the same with or the reverse to each other, and 0 indicates irrelevance. The result (Table \ref{tab:rank-region-correlation}) shows that per capita income is more relevant to predicted group status than per capita GDP. This implies the different nature of the two statistics. Per capita income is a direct measure of people's income, while per capita GDP involves other social and economic factors, which can not be captured by social network data.

\begin{figure}
	\centering
	\begin{subfigure}[b]{1.5in}
		\includegraphics[width=\textwidth]{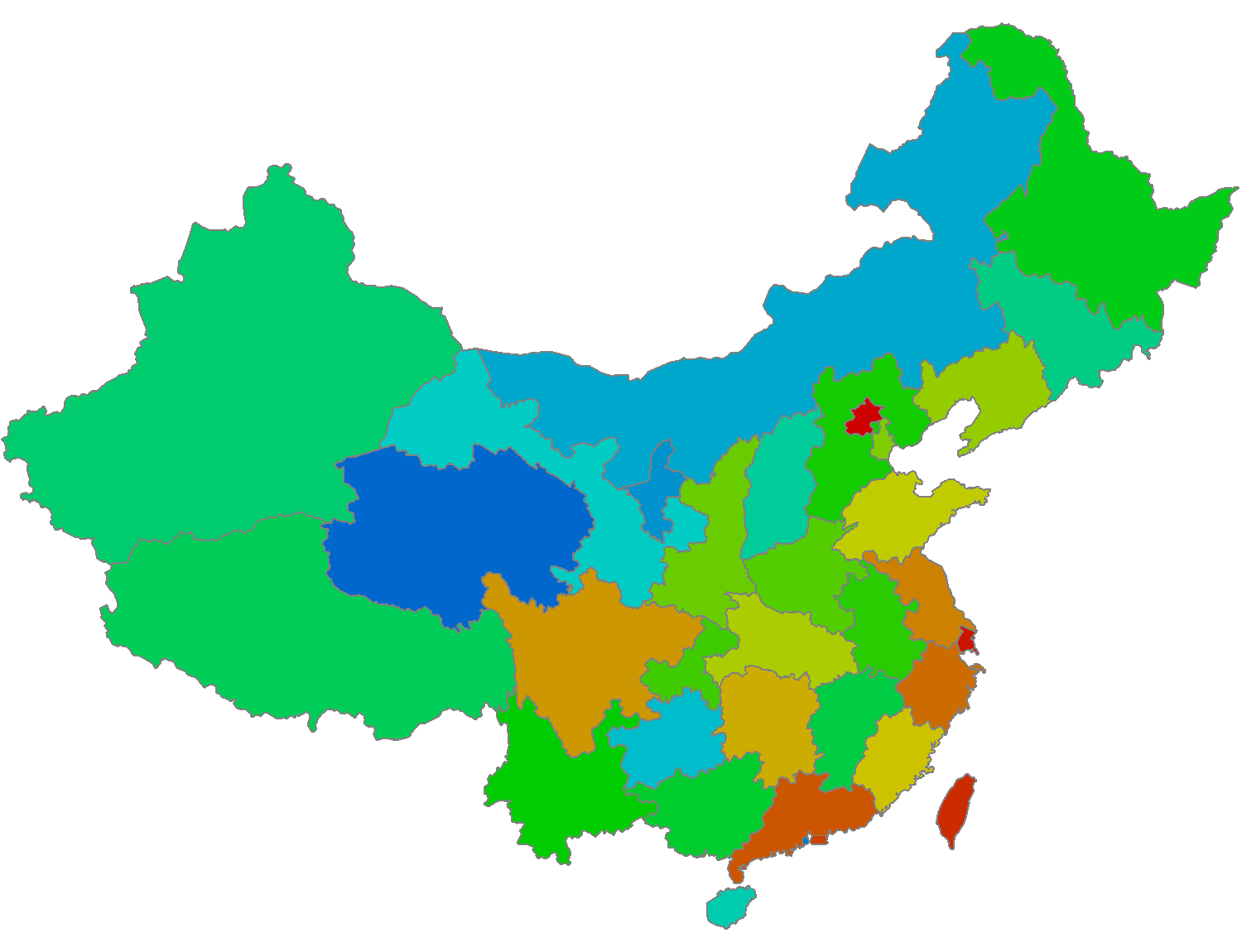}
		\caption{SP}
	\end{subfigure}
	\begin{subfigure}[b]{1.5in}
		\includegraphics[width=\textwidth]{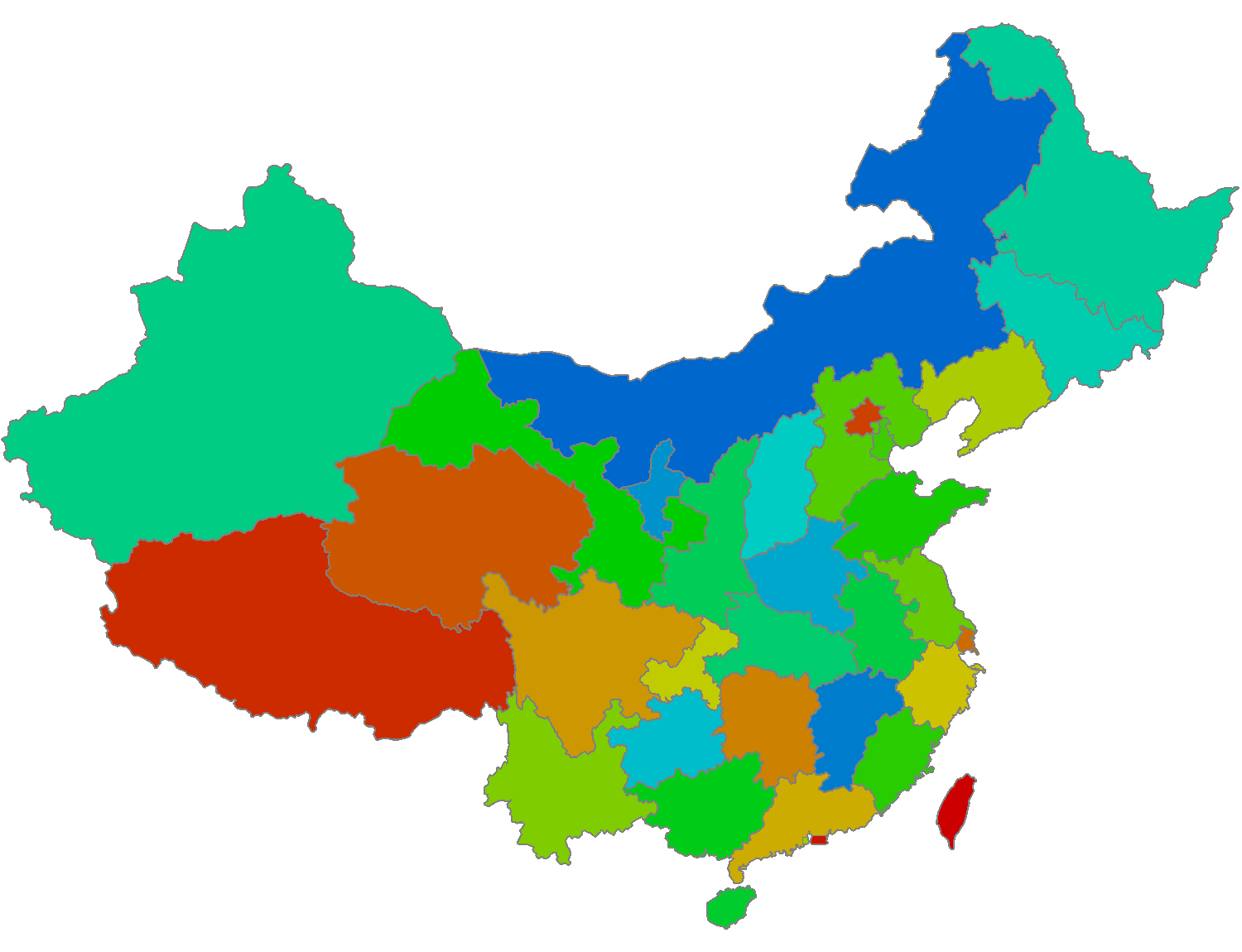}
		\caption{PR}
	\end{subfigure}
	\begin{subfigure}[b]{1.5in}
		\includegraphics[width=\textwidth]{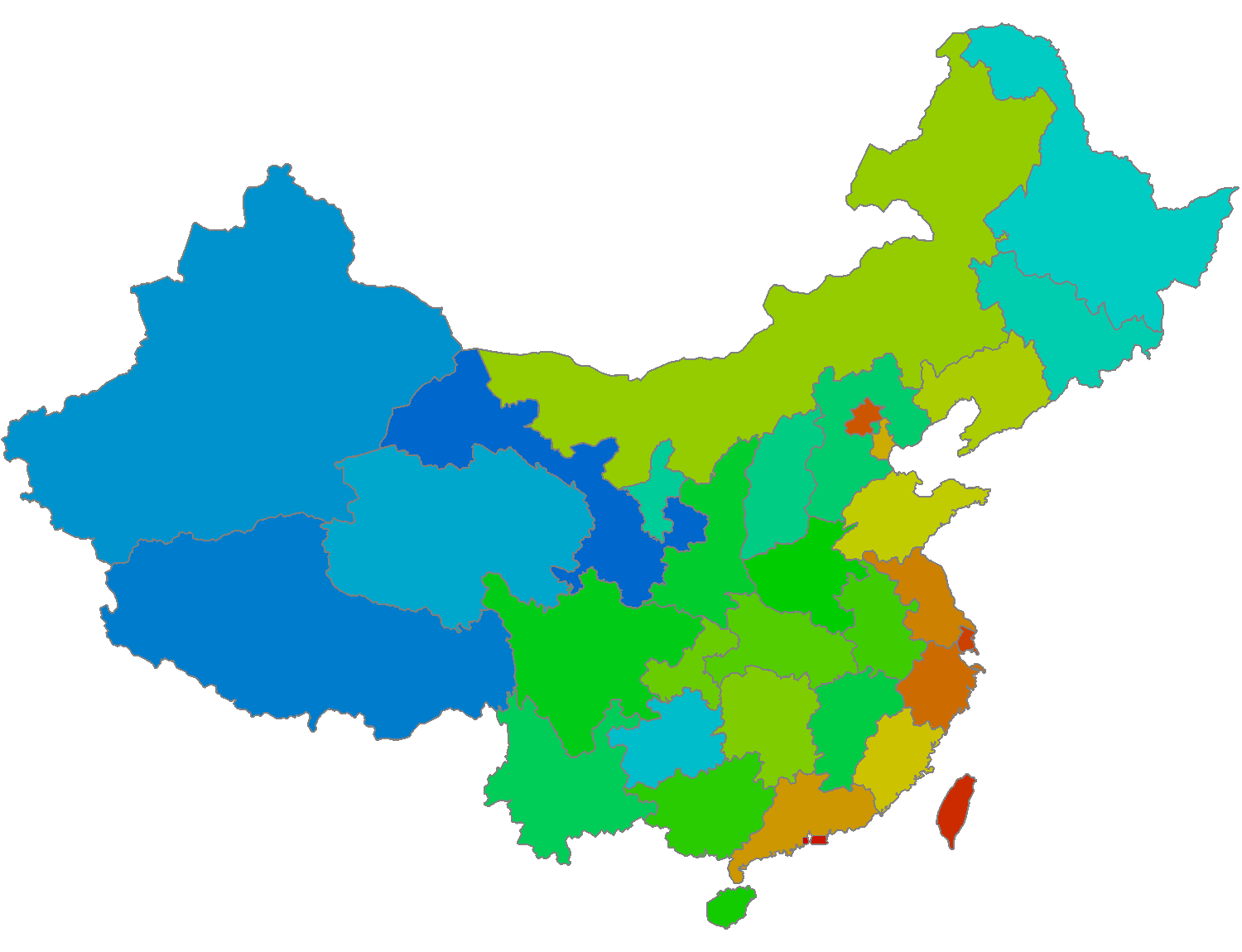}
		\caption{Per capita income}
	\end{subfigure}
	\begin{subfigure}[b]{1.5in}
		\includegraphics[width=\textwidth]{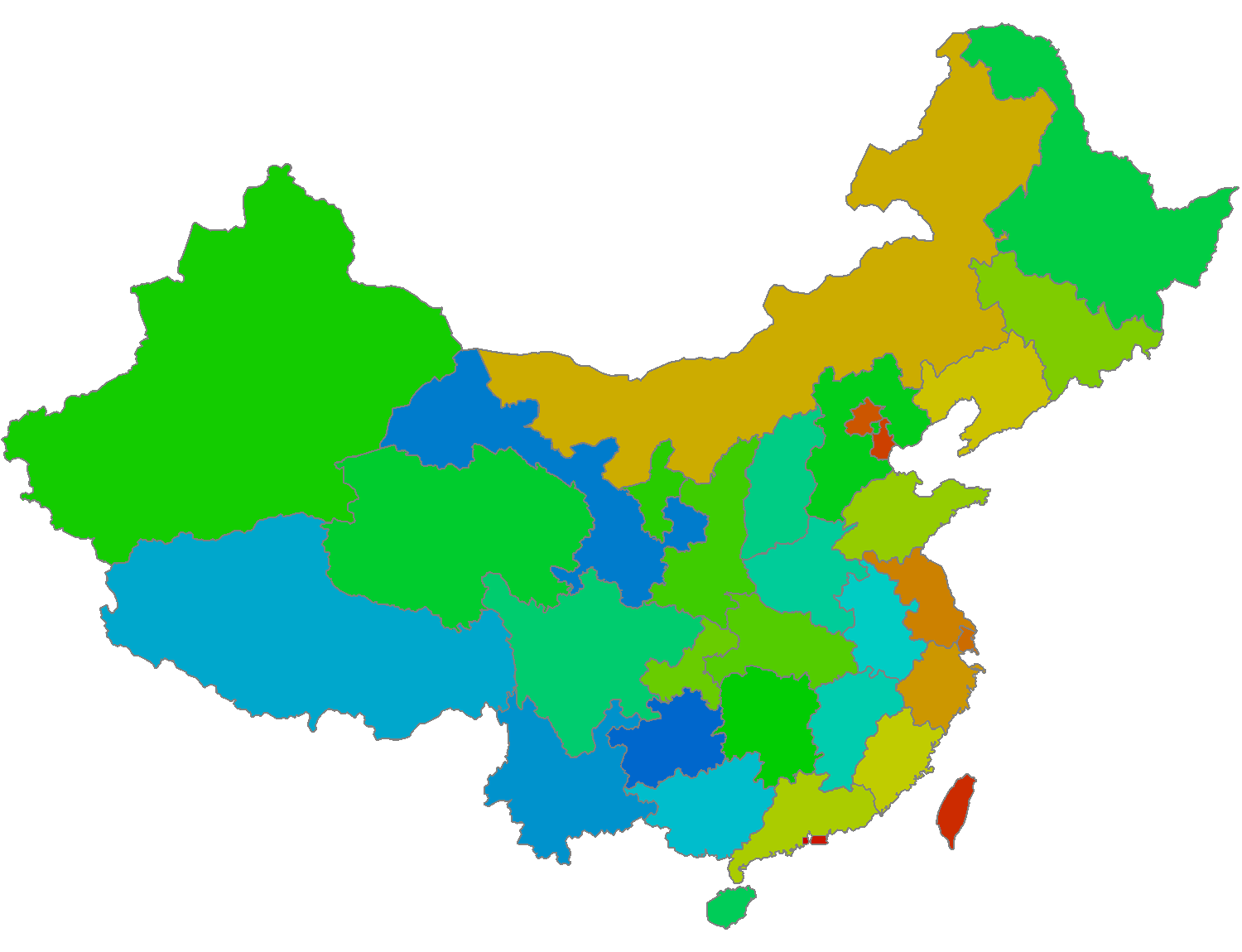}
		\caption{Per capita GDP}
	\end{subfigure}
	\includegraphics{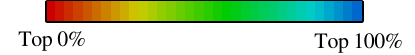}
	\caption{Ranking of regions by group status and socio-economic statistics. UP is similar to SP and thus omitted.}
	\label{fig:rank-region}
\end{figure}

SP and UP produce similar rankings, while UP performs slightly better. PR is significantly less correlated to the statistics than SP and UP. To see the difference, we plot the regions in Figure \ref{fig:rank-region}. Figures \ref{fig:rank-region}(c) and \ref{fig:rank-region}(d) show that regions along the coast in Eastern and Southern China are more developed, and Western China is less developed. The result of SP/UP (Figure \ref{fig:rank-region}(a)) agrees with this tendency.

The PR approach assigns Tibet and Qinghai, which are located in Western China, with unusually high status. In our dataset, 67 and 54 verified users are located in Tibet and Qinghai respectively, which are much less than the average number (2,125). We find 13 out of the 121 users are verified as Tibetan Buddhist leaders. We do not observe such a large fraction of religious leaders in other regions. This can be explained by the fact that Tibetan Buddhism spreads mainly in these two regions nowadays. 7 of the Tibetan Buddhist leaders are among the top 20 users in these two regions. When we discard them in the PR approach, Tibet and Qinghai are ranked at \#17 and \#33 respectively. This shows again the sampling bias in social network data, where verified users may not represent the average population. By inferring unverified users, the bias can be partially fixed.

The top 5 regions in terms of per capita income are Beijing, Shanghai, Hong Kong, Macao, and Taiwan. Four of them are also ranked in the top 5 by SP and UP, but Macao is assigned with surprisingly low status. We find that only 40 verified users are located in Macao. Inference on such a small sample seems unreliable, even if membership inference is applied. A better coverage of users may solve this issue.

\section{Related Work}
\label{sec:related-work}
Social status is one of the fundamental principles that describe the stratification of society. A related concept, namely social capital \cite{putnam1993prosperous}, regards one's social ties as a resource that can be utilized to access more resources \cite{ellison2007benefits}. It is different from social status as the later focuses on the position in society and explains social ties as a consequence of status. A few works have studied social status in social networks. Leskovec et.\ al (\citeyear{leskovec2010}) find that the structure of signed networks can be explained by status theory. Hopcroft et.\ al (\citeyear{hopcroft2011followback}) observe that users with similar social status tend to follow each other reciprocally on Twitter. Many works focus on ranking with different principles, e.g., influence. However, little is known about how online status is related to reality. In this work, we try to answer this question with the proposed method and evaluation on three different domains. Our method is built on the ideas and techniques developed for social network influence analysis and user profiling, so we present a brief discussion of them here. We also distinguish our work with a similar problem, namely ranking in heterogeneous network.

\noindent\textbf{Influence in microblogs}
A number of works focus on analyzing influence in microblogs based on the network structure \cite{kwak10www} and content of tweets \cite{cha2010measuring}. Different models are proposed to measure the influence \cite{bakshy2011,romero2011}. Another body of research focuses on topic-specific influence in microblogs. Given certain topics, the task is to find the most influential users in consideration of relevance and importance \cite{pal2011,ghosh2012cognos,bi2014}.

Some works start with PageRank to design better ranking models \cite{tang2009influence,weng2010wsdm,ghosh2012www}. In our work, we also find it a good measure of social status. We are aware that better models for social status can be developed, and leave it for future work.

\noindent\textbf{Ranking in heterogeneous network}
A social network with membership information can be modeled as a heterogeneous network, containing two types of nodes: users and groups. A user and a group are linked by an edge if the user is a member of the group. Predicting group status is casted to the ranking problem in the heterogeneous network. Sun et al.\ (\citeyear{sun2009netclus}) proposed a family of algorithms that perform clustering and ranking simultaneously. They applied the algorithms on DBLP to cluster and rank academic papers and conferences. Ji et al.\ (\citeyear{ji2011rankclass}) developed methods to rank and classify nodes simultaneously.

The above works assume that links in the network are fully presented. In our scenario, only a small fraction of biased links between groups and users are known, so these algorithms can not work properly. In addition, online social networks, where users can establish links casually, are much more noisy than citation networks or co-author networks.

\noindent\textbf{User profiling}
Membership inference can be considered as a user profiling problem, where missing attributes are inferred from known attributes and network structure. Several works are built on the idea of propagation. Backstrom et al.\ (\citeyear{backstrom2010}) proposed an algorithm specialized to locations. Mislove et al.\ (\citeyear{mislove2010}) infer user profiles in a university by applying a graph clustering algorithm. However, members of the detected cluster do not necessarily share the same attribute. Pennacchiotti and Popescu (\citeyear{pennacchiotti2011}) proposed a hybrid algorithm to infer binary attributes, by combining machine learning methods and propagation on networks. More complicated models are proposed to capture the correlation between different types of attributes \cite{zhao2013}, and the relationship between node and edge \cite{li2014}. Limitations of the user profiling problem were discussed in \cite{cohen2013classifying}.

Given that we are processing a network with millions of nodes and hundreds of groups, we need an inference algorithm that is scalable to both the size of network and the number of groups. We notice that little attention has been paid to social tie strength in user profiling. We leverage the techniques developed for supervised random walk \cite{backstrom2011rw,gao2011srv} to learn the strength. It turns to be reasonably efficient and effective.

\section{Conclusion}
\label{sec:conclusion}

In this paper, we investigate the relationship between social network and social status in reality. Our method consists of two steps, namely membership inference and predicting group status. We compare the predicted status with various indicators from the real world, covering domains of university, occupation, and region. Our results show that it is possible to predict social status via social networks with reasonable accuracy. However, inconsistence between the status in social networks and reality is observed, implying challenges of this problem.

The inconsistence comes in two folds. First, as we use the membership information as a bridge to link social network and the real world, it is unclear what kind of users would provide such information. We find bias in user samples in all three domains. Elite graduates from non-top universities rather than average graduates would have their accounts verified. While officials try to hide their identities, marketers are eager to promote themselves. Religious leaders, who are very prestigious, concentrate in certain regions. Although users can specify whatever profiles they want, they have no direct control on their friends' profiles \cite{mislove2010}. This suggests membership inference as a potential way to fix such bias. We believe that this should not be a big problem if more comprehensive data is available.

Second, users do not behave equally active in social networks. Graduates from art, music, or acting schools are quite popular on social networks. It is due to the occupational nature of the schools. We also find evidence that marketers try to gain influence by link farming, rendering their online status higher than they have in reality. Our current measure of social status does not take these issues into consideration. We consider this as a challenge for future studies, since it is unclear yet if these behaviors are related and if it is possible to describe them in a unified model. 

We believe the study of social status is valuable to several applications beyond the discovery itself. We show that social networks imply social status in reality. Better recommendation systems could be designed for online shopping by considering social status, since it is closely related to income and educational attainment. Social status can also be utilized to analyze trust and influence on social networks. Other types of data could also be investigated, e.g., retweeting/sharing activities, textual contents, and images. The proposed approach may also serve as a tool to analyze social status of other entities, e.g., companies, websites, books, and movies.

%\noindent \textbf{Acknowledgements} We would gratefully thank Professor Yu Xie from University of Michigan and Guoying Huang from Peking University for their informative introduction on social status studies in sociology.

%\setstretch{0.96}
%\let\OLDthebibliography\thebibliography
%\renewcommand\thebibliography[1]{
%	\OLDthebibliography{#1}
%	\setlength{\parskip}{0pt}
%	\setlength{\itemsep}{1.5pt}
%}
\bibliographystyle{aaai}
\bibliography{reference}

\end{document}